\documentclass[11pt]{iopart}

\usepackage[compress,sort]{cite}
\usepackage{verbatim}

\usepackage{subfig}
\usepackage{xcolor}
\usepackage{graphicx}
\usepackage{dcolumn}
\usepackage{bm}
\newcommand{\beq}{\begin{equation}}
\newcommand{\eeq}{\end{equation}}

\newcommand{\vt}[1]{\boldsymbol{#1}}

\DeclareMathAlphabet\mathbfcal{OMS}{cmsy}{b}{n}

\newcommand{\ket}[1]{\left| #1 \right\rangle}
\newcommand{\bra}[1]{\left\langle #1 \right|}

\newcommand{\ketbra}[2]{\left| #1 \rangle \langle #2 \right|}

\begin{document}
\nocite{*}

\title{Nonlinear quantum response of a local surface plasmon coupled to a 2D material}

\author{Daniel B. S. Soh$^{1,2}$, Ryotatsu Yanagimoto$^1$, Eric Chatterjee$^1$, and Hideo Mabuchi$^1$}

\address{$^1$E. L. Ginzton Laboratory, Stanford University, Stanford, California 94305, USA.}
\address{$^2$Sandia National Laboratories, Livermore, California 94550, USA.}

\begin{abstract}
	We present a theoretical study of the optical response of a nonlinear oscillator formed by coupling a metal nanoparticle local surface plasmon resonance to excitonic degrees of freedom in a monolayer transition-metal dichalcogenide. We show that the combined system should exhibit strong anharmonicity in its low-lying states, predicting for example a seven order-of-magnitude increase in nonlinearity relative to a silicon photonic crystal cavity. {\color{blue} Then, we demonstrate that such system exhibits strong quantum features such as antibunching and non-Gaussianity.} Arrays of such nanoscale nonlinear oscillators could be used to realize novel optical metamaterials; alternatively, an individual nanoparticle-monolayer construct could be coupled to an optical resonator to mediate efficient input-output coupling to propagating fields.
\end{abstract}

\ead{dansoh@stanford.edu}
\vspace{10pt}
\begin{indented}
	\item[]April 2019
\end{indented}

\section{Introduction}

{\color{blue} The quest for realizing a nanophotonic nonlinear oscillator has remained as a formidable challenge up until now, despite that many discussions on how Kerr optical nonlinearity can be used to realize such a nonlinear oscillator \cite{mabuchi2012qubit,imamoglu1997strongly}. The hardship is closely linked to another equally difficult quest for nanophotonic single-photon optical nonlinearity. The optical nonlinear oscillator, if realized, will be a game changer as it would operate at GHz or even at THz in a scalable platform. It is notable that many materials with a large Kerr coefficient have been reported so far, and yet, none of these succeeded to realize such a nonlinear oscillator. As we will explain in this paper, merely having a large Kerr nonlinearity is not sufficient since those materials with large Kerr nonlinearity tend to have also strong two-photon absorption, which forestalls the nonlinear response.} 

Atomically thin 2D materials are known to possess enormous optical nonlinearities within subatomic thickness \cite{sun2016optical,autere2018nonlinear,soh2016comprehensive,soh2018optical}. However, such a small thickness of the 2D materials is usually viewed as a drawback as their incorporation in conventional optical resonators provides disappointing nonlinear-optical performance because of small filling fraction in wavelength-scale volumes. For example, the excellent review papers by Sun \textit{et al.\ }\cite{sun2016optical} and Autere \textit{et al.\ }\cite{autere2018nonlinear} illustrate the integrated optical modulation capability of 2D materials. However, the authors of these papers clearly pointed out that, to fully harness the potential of the strong nonlinearity of 2D materials at the atomic scale, it is necessary to circumvent the issues of inefficient light-matter interaction due to 2D material’s subnanometer thickness. 

In order to accomplish a large optical nonlinearity in a small volume, various approaches have been extensively studied to date. For example, Nielsen \textit{et al.\ }studied the plasmonic nanofocus driving strong four-wave mixing~\cite{nielsen2017giant}. However, the authors acknowledged that their accomplished optical nonlinearity is far below the required level for quantum effects. 

We propose a novel approach to harnessing the extreme field confinement of localized surface plasmon resonances (LSPR) for nonlinear optics, using 2D monolayer materials. {\color{blue}This approach is a unique combination of the freedom to adjust the real and the imaginary values of the third-order Kerr nonlinearity from the excitons of 2D materials and the field enhancement from plasmonic coupling.}  Our result clearly demonstrates an evident quantum phenomenon – the strong anti-bunching effect. Thus, our result is a crucial step towards the long-standing quest for the strong quantum effect of nano-structured materials.  

A LSPR of a metal nanoparticle (MNP) may be modeled as a quantized harmonic oscillator~\cite{WAKS2010CAVITY}. For a given resonance frequency, the energy states form a harmonic ladder broadened by Ohmic loss of the metal~\cite{shvets2012plasmonics}. The LSPR field is confined within a volume slightly larger than the MNP, which is significantly smaller than the free-space wavelength of light at the LSPR resonant frequency~\cite{jensen2000nanosphere}. We analyze a scheme for turning an LSPR into a quantum {\em nonlinear} resonator via near-field coupling to an atomically thin 2D material, such as a transition metal dichalcogenide (TMD). Two key considerations motivate this proposal: that monolayer TMDs can provide strong Kerr nonlinearity with a favorable coherence-to-decoherence ratio \cite{soh2018optical}, and that monolayer TMDs are sufficiently thin to overlap substantially with the nanoscale LSPR evanescent field. Then, we proceed to demonstrate the remarkable quantum features of such systems, namely, photon antibunching and non-Gaussian state generation. 

{ \color{blue} The structure of this article is as follows: section 2 builds up the concept of nonlinear oscillators, section 3 describes the prescription for highly nonlinear oscillators, section 4 models the quantum dynamics of the nonlinear oscillator and presents the simulation results, and finally, a conclusion and discussions follow.}

\section{Nonlinear oscillator}

There are many complementary schemes for inducing nonlinearity in an optical oscillator~\cite{imamoglu1997strongly,mabuchi2012qubit,miranowicz2013two}. In the (arguably) most widely applicable approach, a Kerr-nonlinear medium can be introduced into an otherwise passive optical cavity. If the ratio of nonlinear Kerr effect to absorption loss is sufficiently large, such a system may be useful for quantum optics and quantum information processing. Unfortunately, it is difficult to find materials with a large nonlinearity-to-loss ratio, which are amenable to incorporation within conventional or nanofabricated optical resonators. For example, although some bulk materials such as silicon have a significant Kerr optical nonlinearity at near-infrared optical frequencies, they also have a large imaginary nonlinear response (two-photon absorption) that gives rise to comparably large losses~\cite{lin2007dispersion}. Setting aside the issue of losses, nanophotonic resonators incorporating conventional optical materials such as silicon generally do not possess a sufficiently large Kerr nonlinearity at low photon numbers for quantum information processing applications.

\begin{figure}[tb]
	\centering
	\vbox{
		\includegraphics[width=0.7\textwidth]{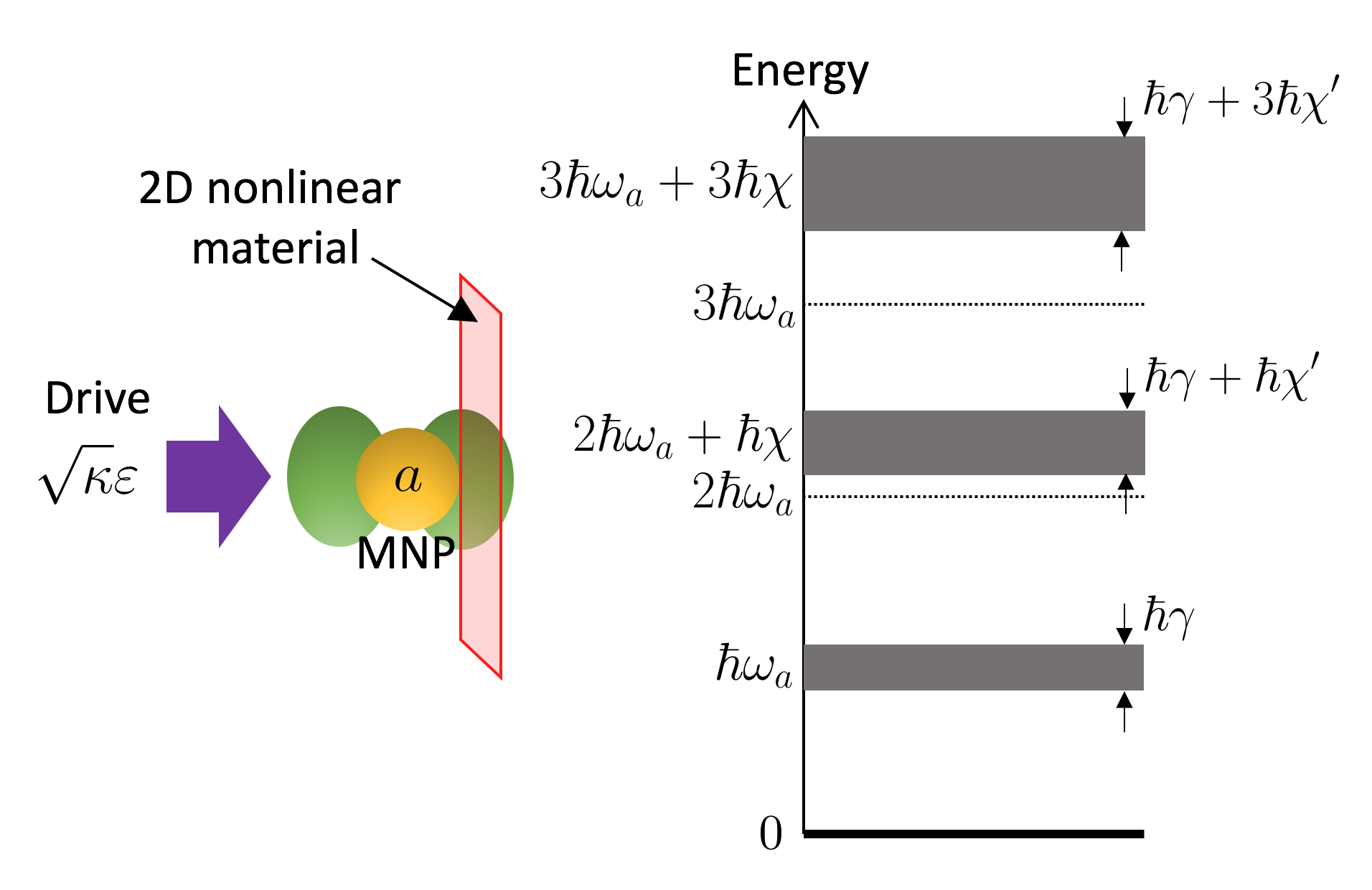}
		\caption{Left: schematic of a metal nanoparticle (MNP) and 2D material. Right: energy levels of the coupled system.}
		\label{fig:schematic}
		\vspace{-0.15in}
	}
\end{figure}

Monolayer MoS$_2$ has an order-of-magnitude larger real $\chi^{(3)}$ than bulk silicon and can have a significant real-to-imaginary ratio of $\chi^{(3)}$ \cite{soh2018optical}.  Its 2D structure furthermore makes it a promising candidate to induce a large coherent nonlinearity through direct coupling to the strongly confined field of an LSPR. Such an LSPR-TMD system could be efficiently coupled to optical input-output modes by incorporation within an optical cavity, in a manner analogous to the way that a Josephson-Junction nonlinear LC oscillator can be coupled to propagating fields by a microwave resonator. With regard to the potential scalability of such a device concept, we note that, while automated assembly of the type of system we envision could be challenging to realize, the material components we require may be producible in bulk by chemical MNP synthesis~\cite{jana2001wet} and chemical-vapor-deposition (CVD) growth of TMDs~\cite{novoselov20162d,rogers2018laser}. One can alternatively envision a large number of MNPs dispersed on a TMD membrane to form an array of nonlinear oscillators with complex couplings in both real and frequency spaces; such an optical metamaterial could provide rich nonlinear dynamics for reservoir computing-type architectures~\cite{lukovsevivcius2009reservoir}.

Let us consider an MNP driven by a classical light source. A monolayer TMD is positioned nearby the MNP so that the LSPR field has a considerably large overlap with the monolayer TMD (see Fig.~\ref{fig:schematic}). To understand the system, we adopt the quantum master equation \cite{carmichael2009statistical,breuer2002theory}, which is particularly suitable for optical frequencies. In this framework, the unitary evolution of the system is modeled through a Hamiltonian $H = H_a + H_\mathrm{drive}$ where \cite{drummond2014quantum,WAKS2010CAVITY,peng2017enhancing,imamoglu1997strongly,ferretti2012single,miranowicz2013two,li2017cascaded} (see \ref{sec:Ham}):
\begin{eqnarray}
H_a &= \hbar \omega_a a^\dag a + \frac{\hbar\chi}{2} a^\dag a^\dag a a, \label{eq:H1}\\
H_\mathrm{drive} &= - i \hbar \sqrt{\kappa} (a^\dagger \varepsilon e^{- i \omega t} - a \varepsilon^* e^{i \omega t}), \label{eq:H2}
\end{eqnarray}
where $\omega_a$ is the resonant frequency of the LSPR field, $\chi$ is the Kerr nonlinear coefficient, and $a, a^\dag$ are the annihilation and the creation operators of the LSPR field. Here, $\kappa$ is the coupling coefficient, $\varepsilon$ is the drive, and $\omega$ is the frequency of the drive field. The nonunitary evolution of the system undergoes dissipation through the Lindblad operators $L_1 = \sqrt{\gamma} a$ with $\gamma = \gamma_r + \gamma_o$, which includes both the radiative ($\gamma_r$) and nonradiative ($\gamma_o$) decays where the nonradiative Ohmic loss is dominant ($\gamma \simeq \gamma_o \gg \gamma_r$)  \cite{shvets2012plasmonics}, and $L_2 = \sqrt{(\chi'/2)} a a$, which is the nonlinear loss (two-photon absorption) through the coupling with the nonlinear 2D material. As a consequence, the effective energy levels appear as in Fig.~\ref{fig:schematic}. Without nonlinear coupling to the 2D material, the energy levels for a given frequency $\omega_a$ are a linear harmonic ladder with an equal level spacing of $\hbar \omega_a$, with a level broadening $\hbar \gamma$. The nonlinearity modifies the energy levels in two ways: both the energy level shift and the level broadening vary with LSPR photon number, resulting in an anharmonic energy ladder. Assuming a single-frequency optical driving field, it is apparent that, in order not to excite an effectively linear (trivial) response, the nonlinear energy shifts must be larger than the level broadenings. 

The nonlinear energy shift and broadening depend on the coefficients $\chi$ and $\chi'$ (both real-valued), which are calculated using the results in \cite{drummond2014quantum,peng2017enhancing,ullah2018analysis,cappellini2001optical} (see the detailed derivation in \ref{sec:chi}):
\begin{equation}
\chi + i \chi' = \frac{27 \epsilon_0 \hbar \omega_a^2}{4} \int d^3 r \; \chi^{(3)} (\vt{r}) | \vt{f}_\parallel (\vt{r})|^4, \label{eq:chi}
\end{equation}
where $\epsilon_0$ is the vacuum permittivity and $\chi^{(3)}(\vt{r})$ is the position ($\vt{r}$) dependent third-order nonlinear susceptibility. Since monolayer TMDs have a negligibly small optical response to out-of-plane-polarized electric fields~\cite{wang2017plane,echeverry2016splitting}, we count only the in-plane LSPR field component (with respect to the monolayer 2D material): $\vt{f}_\parallel (\vt{r})$ is the in-plane component of the LSPR mode function $\vt{f} (\vt{r})$, normalized as $\int d^3 r ~ \epsilon_0 \epsilon_r (\vt{r}) |\vt{f}(\vt{r})|^2 = 1$ with the position dependent relative permittivity $\epsilon_r (\vt{r})$,  Evidently, the overlap between the LSPR mode field and the 2D material volume must be large to increase $\chi$. We note that a large overlap also increases the nonlinear broadening of two-photon absorption. Hence, the the ratio $\chi/\chi'$ must be large, which depends in general on the material and the operating frequency.

To appreciate how difficult it is to make a practical nonlinear (anharmonic) oscillator, let us consider an optical resonator filled with silicon. We make an example of the state-of-the-art smallest photonic crystal cavity to increase the nonlinear energy shift: a cavity of volume $V_c = 0.02$  $\mu$m$^3$ having a cavity $Q = 2 \times 10^5$ \cite{kuramochi2017ultrahigh}. Then, using $\mathrm{Re}[\chi^{(3)}] \sim 4.8 \times 10^{-20}$ m$^2$/V$^2$ of silicon at 1 eV photon frequency (energy) \cite{lin2007dispersion}, \Eref{eq:chi} leads to $\hbar \chi \sim 1 \times 10^{-9}$ eV. However, the linear cavity-lifetime-induced level broadening (ignoring the intrinsic linear loss of silicon) is $\hbar \gamma \sim 5 \times 10^{-6}$ eV. Hence, the line broadening is dominant over the nonlinear energy level shift at low photon number. Although quantum well materials such as GaAs/InGaAs have appreciable Kerr nonlinearity ($\mathrm{Re}[\chi^{(3)}] = 1.3 \times 10^{-19}$ m$^2$/V$^2$ at 1.3 $\mu$m wavelength \cite{hales2018third}), the two-photon absorption is large ($\mathrm{Im}[\chi^{(3)}] = 1.4 \times 10^{-18}$ m$^2$/V$^2$ \cite{hales2018third}) and the linear absorption is excessively large ($\alpha > 10$ ~cm$^{-1}$) \cite{akiyama2001nonlinearity,yoshida1998intersubband} so that the nonlinearity-to-broadening ratio is even less favourable than our silicon example. Chalcogenide glass materials are not much better due to the same reason of large linear and two-photon absorption~\cite{zakery2003optical}.

\section{Localized surface plasmon coupled to 2D materials}

We will show below that our proposed system could achieve few-photon nonlinear energy shifts that dominate broadening, and thus realize a quantum nonlinear oscillator. We first discuss several design aspects for the system. We utilize the large ratio of the real and the imaginary values of $\chi^{(3)}$ in a monolayer MoS$_2$ by detuning slightly from the two-photon resonant frequency 1.06 eV~\cite{soh2018optical}. We must also design an MNP to support an LSPR resonance near 1.06 eV with a sufficiently small $\hbar \gamma$. The resonance frequency $\omega_a$ of the LSPR mode is well known to depend upon the geometry of the MNP \cite{link2000shape,kelly2003optical,wang2006general}, with thin MNPs exhibiting lower resonance frequencies~\cite{kirakosyan2016surface}. The reason is that the portion of the LSPR field inside the metal reduces as the metal layer becomes thinner, which tends to red-shift the resonance frequencies~\cite{wang2006general}.

We need to maximize the in-plane component of the LSPR field that overlaps the monolayer MoS$_2$. A natural choice of geometry is a disk shape MNP with a large ratio of diameter to thickness. A large diameter improves the in-plane LSPR field component while a thinner disk shape pulls $\omega_a$ down from the nominally visible plasmonic resonances of silver. Using finite-element-method (FEM) software (COMSOL), we found that a diameter-to-thickness ratio of 22 for a silver disk should provide an LSPR resonance frequency near the target. On the other hand, the value of $\gamma$ of an LSPR is known to be independent of shape and size, since it is completely determined by the complex dielectric function of the material; once the material and operating frequency are determined, there is not much one can do to adjust $\gamma$ \cite{wang2006general}. For this theoretical study, we take from the work of Wang and Shen a Q-factor $Q = \omega_a/\gamma = 50$~\cite{wang2006general}. In reality, the surface imperfections may reduce the Q-factor.

We also need to consider potential impacts on the intrinsic optical properties of a TMD monolayer when an MNP is placed nearby---an excessively small gap between the two may alter the band structure of the 2D material and impact the nonlinearity adversely. To address this problem, we consider a tight binding band Hamiltonian $H_\mathrm{band} = \sum_{<i,j>} J_{ij} c_i^\dag c_j + \sum_i L_i c_i^\dag c_i + \sum_i M_i (c_i^\dag d + c_i d^\dag)$ where $c_i^\dag, d^\dag$ are the second-quantized fermionic creation operators for the electrons in the 2D material lattice and the impurity nearby, respectively \cite{marder2010condensed}. The first term is the nearest-neighbor (nn) hopping energy, the second is the on-site energy, and the third is the interaction energy between the 2D material and the impurity sites. We note that $J_{ij}$ and $M_i$ originate from the Coulomb potential. It is customary to ignore the next-nearest neighborhood (nnn) hopping since the matrix element for this remote potential is negligibly small compared to the nearest-neighbor hopping. Hence, to preserve the band structure intact, we require the condition $|M_i| \ll |J_{ij}|$, which can be fulfilled if the distance between the 2D lattice sites and the impurity sites becomes larger than the next-nearest neighbor (nnn) distance of the 2D lattice. When the hexagonal atomic arrangement of monolayer TMDs is considered, the nnn distance is 0.32 nm (MoS$_2$), approximately 1.7 times longer than nn. Indeed, Liu \textit{et al.\ }recently discussed the encapsulation of monolayer TMDs and found that using a few-layer hexagonal boron-nitride (hBN) buffer preserved the properties of monolayers successfully \cite{liu2016van}. Man \textit{et al.\ }discussed the successful protection of monolayer properties using even a single layer hBN buffer~\cite{man2016protecting}. Hence, in this theoretical paper, we set the minimum distance between the monolayer TMD and the MNP as 0.3 nm, corresponding to the thickness of a single layer hBN~\cite{golla2013optical}. Additionally, since the operating frequency is detuned from both the one- and two-photon excitonic resonance of the monolayer TMD, exciton generation will be minimal and we neglect metal-dielectric effects such as mirror charges.

\section{Dynamical model and simulation results}

To understand the dynamics of the system, we consider the adjoint master equation for a Heisenberg picture operator $Q$ as \cite{breuer2002theory}:
\begin{equation}
\dot{Q} = - \frac{i}{\hbar} [Q,H] + \sum_j \left( L_j^\dag Q L_j - \frac{1}{2} L_j^\dag L_j Q - \frac{1}{2} Q L_j^\dag L_j \right).
\end{equation}
Then, we easily obtain the dynamical equation for the LSPR field using the Hamiltonian in \Eref{eq:H1} and \eref{eq:H2} and the Lindblad operators:
\begin{equation}
\dot{a} = - i \Delta_a a - i \chi (a^\dag a) a - \frac{\gamma}{2} a - \frac{\chi'}{2} (a^\dag a) a - \sqrt{\kappa} \varepsilon, \label{eq:a}
\end{equation}
where $\Delta_a = \omega_a - \omega$. In the limit of considering only the lowest three energy levels with a moderate drive power, the steady-state population in the second excited level can be expressed using the population in the first level as (\ref{sec:pop}):
\begin{equation}
\rho_{22,\mathrm{ss}} = \frac{4 \kappa \varepsilon^2/(\gamma + \chi')^2}{1 + (4 \chi^2 + 4 \kappa \varepsilon^2)/(\gamma+\chi')^2} \rho_{11, \mathrm{ss}}. \label{eq:rho22}
\end{equation}
Note that for a linear system ($\chi=\chi'=0$), the population is $\rho_{22,\mathrm{ss}} = \rho_{11,\mathrm{ss}} (4 \kappa \varepsilon^2/\gamma^2)/(1 + 4 \kappa \varepsilon^2/\gamma^2)$, which exhibits the conventional saturation feature with respect to the drive power.  The nonlinearity (anharmonicity) kicks in if the quantity $4 \chi^2/(\gamma+\chi')^2$ is sufficiently large to suppress $\rho_{22,\mathrm{ss}}$. Hence, we should adjust the system design parameters to maximize the nonlinear figure of merit (FOM) $4 \chi^2 / (\gamma + \chi')^2$. 

The FOM depends mainly on two design parameters: the MNP size and the drive frequency detuning. The MNP size determines the overlap between the monolayer TMD and the LSPR field, scaling $\chi$ and $\chi'$ in the same fashion. A large MNP has a more delocalized field outside the MNP and, thus, the relative portion overlapping the thin nonlinear 2D material is small. On the other hand, detuning of the drive frequency adjusts the ratio between $
\chi$ and $\chi'$. The values $\mathrm{Im}[\chi^{(3)}]$ of a monolayer MoS$_2$ roughly follows a typical Lorentzian lineshape while $\mathrm{Re}[\chi^{(3)}]$ is the frequency-derivative of $\mathrm{Im}[\chi^{(3)}]$ \cite{soh2018optical}. Hence, FOM is expected to be nearly zero at the two-photon resonance of the monolayer TMD exciton while the maximum occurs an appropriately detuned frequency. \ref{sec:design} presents the detailed design considerations.

\begin{figure}[tb]
	\vbox{
		\centering
		\includegraphics[width=0.6\textwidth]{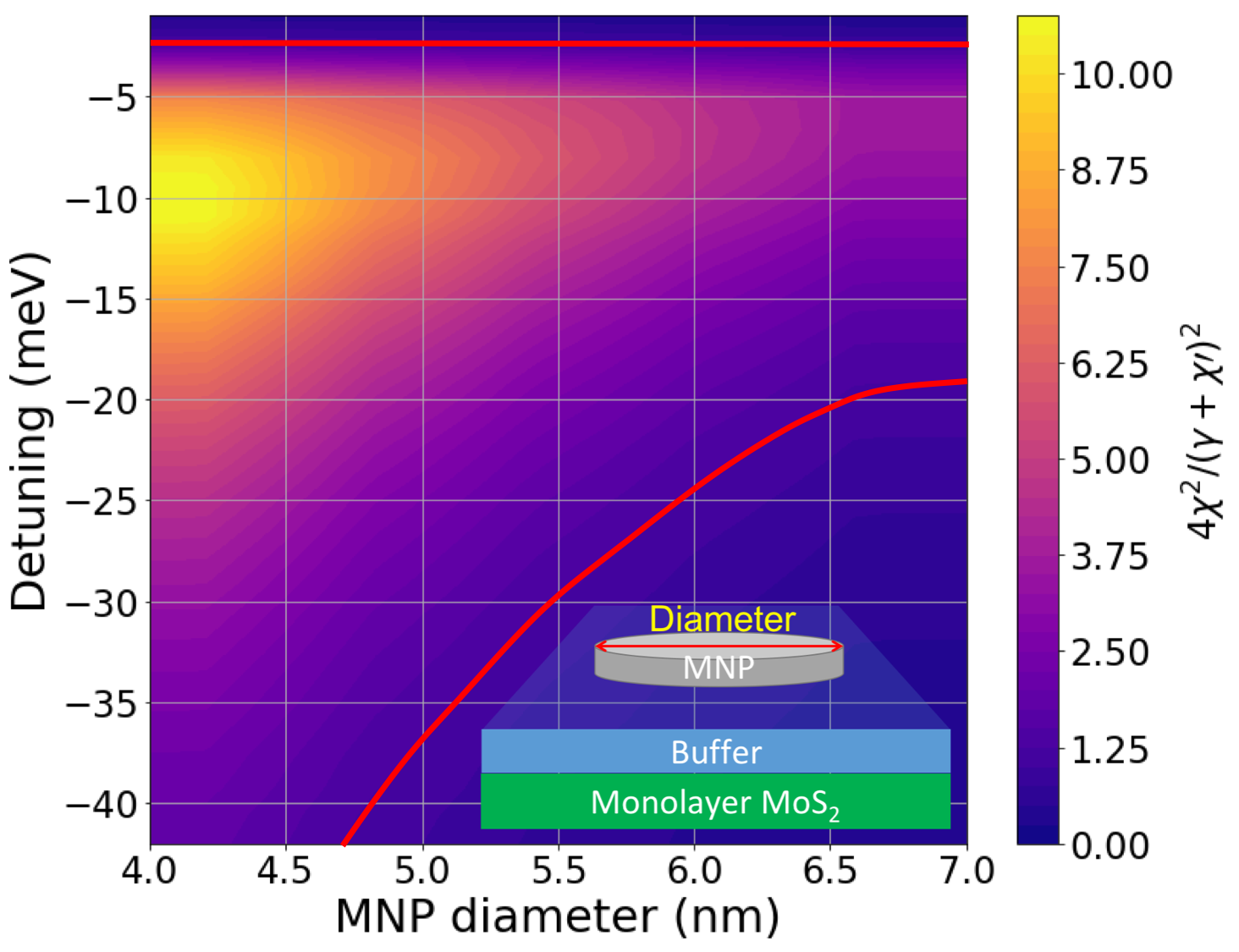}
		\caption{Nonlinear FOM dependence on the diameter of a silver disk MNP and the drive frequency detuning. Solid red lines represent the points with unity FOM (onset of nonlinearity). Inset: geometry of MNP-TMD nonlinear system (not in scale, only the lower layer of monolayer MoS$_2$ is shown for clarity).}
		\label{fig:opt_FOM}
		\vspace{-0.15in}
	}
\end{figure}

To maximize the overlap, we place two monolayer MoS$_2$ planes sandwiching the disk-shaped MNP, with hBN buffers separating the MNP from each of the monolayers. Using an FEM software package (COMSOL), we calculate the detailed field distribution of a given MNP design and subsequently determine $\chi, \chi'$ according to \Eref{eq:chi}. Fig.~\ref{fig:opt_FOM} shows the calculated FOM as a function of the silver disk MNP's diameter (while fixing the diameter-to-thickness ratio at 22) and the drive frequency detuning from the monolayer MoS$_2$'s two-photon resonance at 1.06 eV. We note that with the fixed diameter-to-thickness ratio of 22, the MNP diameter must be larger than 4 nm so that the MNP disk thickness is larger than the atomic size of silver. Remarkably the nonlinear FOM is larger than unity over a wide range of parameters; the unity FOM value is the on-set for a system to be nonlinear. The maximum FOM of 10.7 occurs at a detuning of $- 11$ meV with the smallest MNP diameter of 4 nm. The real and the imaginary values of $\chi^{(3)}$ of the monolayer MoS$_2$ at this detuning are $4.0 \times 10^{-19}$ m$^2$/V$^2$ and $1.0 \times 10^{-19}$ m$^2$/V$^2$, respectively \cite{soh2018optical}. The subsequent $\hbar\chi$ and $\hbar \chi'$ values are obtained as 53 meV and 13 meV, respectively. It is noteworthy that the improvement of FOM by placing additional stacks of monolayer  MoS$_2$ with gap-buffers is negligible due to the fast decaying LSPR mode field. 

{ \color{blue} We should mention that a nanodisk with 4 nm diameter with an aspect ratio of 22 approaches the material limit of a single atom layer silver disk. In this limit, one might suspect the validity of the classical approach (COMSOL) as well as manufacturability of such thin nanoparticles. However, several experimental results on 2D single-layer silver and gold nanoparticles for bio-imaging applications were reported \cite{usukura2014highly,tanaka2015characteristics,masuda2017high}. The 2D silver and gold sheets were fabricated using a self-assembly technique on a water-air interface. Such top-down fabrication approach may place the nanoparticle precisely in the desired location, but the distance among nanoparticles is challenging to be smaller than 100 nm, while other bottom-up (growth) fabrication technique may randomly place the sheet nanoparticles, but with much higher nanoparticle density \cite{usukura2014highly}. Moreover, a good agreement between experimental results and theoretical analysis based on classical Maxwell equations was reported \cite{usukura2014highly,tanaka2015characteristics}. This implies that the theoretical extrapolation towards such thin nanoparticle geometry based on the classical approach may still be acceptable. Moreover, the FOM has a long tail for larger diameter nanodisks. For example, a nanodisk with 10 nm diameter that has $\sim$ 0.5 nm thickness still has FOM of 2.3. We note that even larger (thicker) nanodisks still maintain FOM values larger than unity.}

\begin{figure}[tb]
		\centering
		\captionsetup[subfloat]{oneside,margin={1.5cm,0cm}}
		\subfloat[Linear system]{\includegraphics[width=0.33\textwidth]{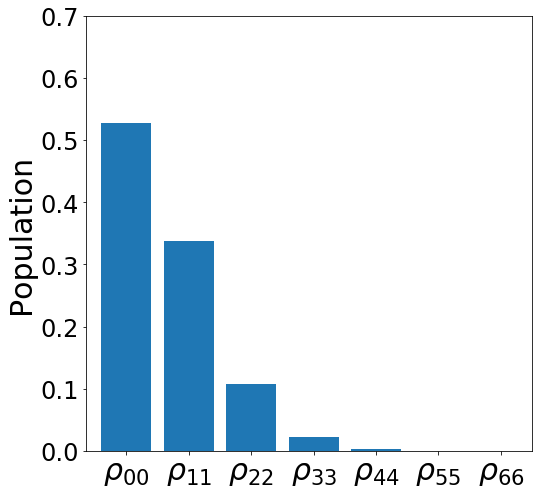}}
		\captionsetup[subfloat]{oneside,margin={0.7cm,0cm}}
		\subfloat[Nonlinear system ($4$ nm)]{\includegraphics[width=0.33\textwidth]{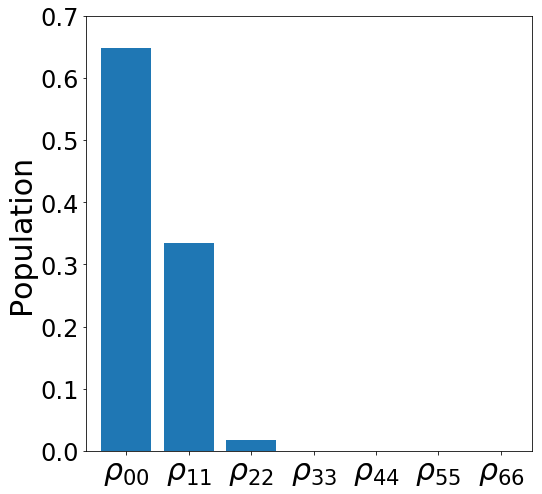}}
		\subfloat[Nonlinear system ($10$ nm)]{\includegraphics[width=0.33\textwidth]{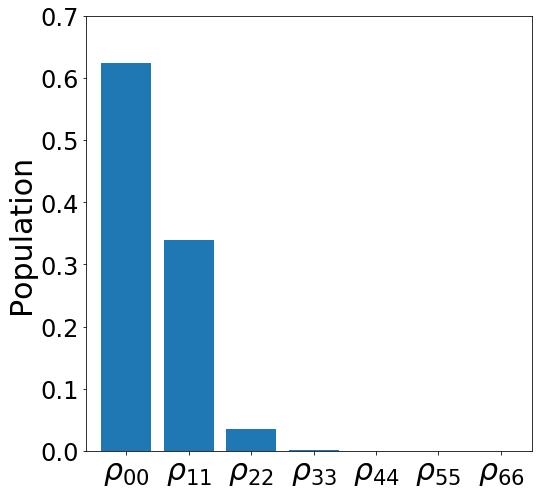}}
		\caption{Comparison of the steady-state excited level populations when driven by an external field at resonance between a linear and a nonlinear system. The middle nonlinear system is from a 4 nm diameter nanodisk ($\hbar \chi = 53$ meV, $\hbar \chi' = 13$ meV) while the right is from a 10 nm diameter nanodisk ($\hbar \chi = 28$ meV, $\hbar \chi' = 16$ meV).}
		\label{fig:population}

\end{figure}

Nonlinear behavior of the system can be clearly seen through the population distribution among energy eigenstates when driven by an external field. For this, we solved the dynamic \Eref{eq:a} by setting $\sqrt{\kappa} \varepsilon = 10$ meV driving on resonance ($\omega=\omega_a$). The steady-state population distributions are shown in Fig.~\ref{fig:population}. The linear system ($\hbar\chi = \hbar\chi' = 0$ meV) shows a typical coherent state population distribution, while the optimal nonlinear system (4 nm diameter, $\hbar\chi, \hbar\chi', \hbar\gamma = 53, 13, 21$ meV) has suppressed population in the upper levels. { \color{blue} Even the simulation with a larger nanodisk (10 nm diameter, $\hbar \chi, \hbar \chi', \hbar\gamma = 28, 16, 21$ meV) shows an excellent suppression of the higher energy level populations, although, for this case, $\rho_{33}$ is nonzero.} The population suppression agrees with the above analysis (\Eref{eq:rho22}). We adjusted the drive power slightly to normalize $\rho_{11}$ since the same drive power will lead to slightly different $\rho_{11}$ populations among the three systems. The residual population $\rho_{22}$ in the nonlinear systems is caused by the saturation of $\rho_{11}$, which could be effectively further reduced using a smaller drive power. 

\begin{figure}[!tb]
	\centering
	\captionsetup[subfloat]{oneside,margin={2.5cm,0cm}}
	\subfloat[4 nm nanodisk]{\includegraphics[width=0.45\textwidth]{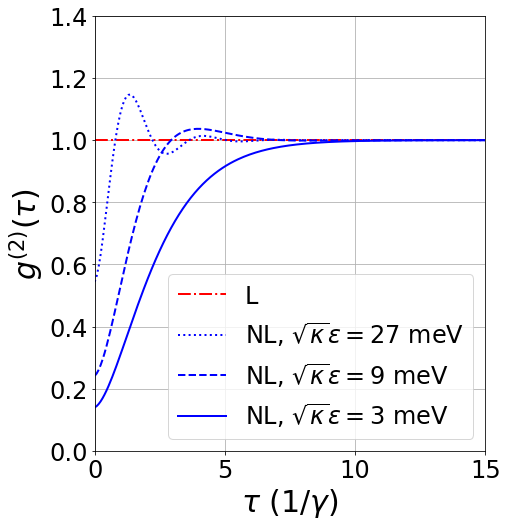}}
	\subfloat[10 nm nanodisk]{\includegraphics[width=0.45\textwidth]{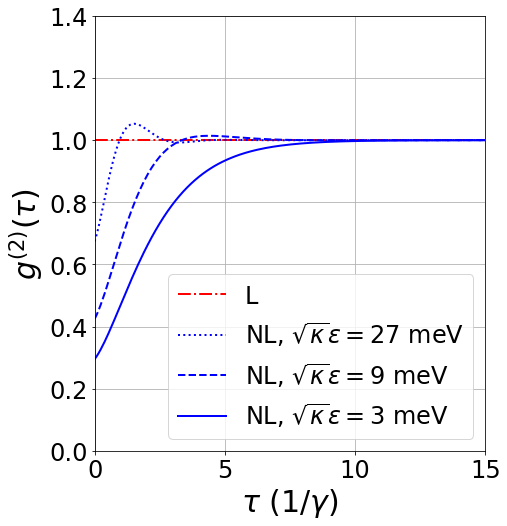}}
	\caption{Two-photon correlation $g^{(2)}(\tau)$ of linear (L) and nonlinear systems (NL) with various drive powers. Left: a 4 nm nanodisk ($\hbar \chi = 53$ meV, $\hbar \chi' = 13$ meV). Right: a 10 nm nanodisk ($\hbar \chi = 28$ meV, $\hbar \chi' = 16$ meV). }
	\label{fig:g2}
\end{figure}

The quantum behavior of the nonlinear system can be clearly seen through the two-photon correlation $g^{(2)}(\tau) = \lim_{t\rightarrow \infty} \langle a^\dagger (t) a^\dagger (t+\tau) a (t + \tau) a(t) \rangle / (\langle a^\dagger (t) a (t) \rangle \langle a^\dagger (t + \tau) a (t + \tau) \rangle)$ \cite{walls2007quantum}. A linear system such as an empty optical cavity will produce $g^{(2)}(\tau) = 1$ for all $\tau$ when driven by a classical source, which corresponds to a coherent state. However, a strongly coupled atom--cavity system exhibits photon blockade and, thus, $g^{(2)}(0)$ close to zero~\cite{walls2007quantum,kimble1992structure}. We obtained the two-photon correlation $g^{(2)}(\tau)$ as shown in Fig.~\ref{fig:g2} by solving the dynamic equation (\Eref{eq:a}). As expected, a linear system ($\chi = \chi' = 0$) shows a flat $g^{(2)}(\tau) = 1$ for all $\tau$, regardless of the drive power. In contrast, the nonlinear system (4 nm nanodisk) shows a strong nonlinearity having $g^{(2)}(0)<1$. Obviously $g^{(2)}(0)$ depends on the population in the upper excited levels and, hence, it depends on the drive power. When weakly driven, the $g^{(2)}(0)$ value reaches as low as 0.15, predicting strong antibunching in optical fields escaping from the LSPR. As the drive power increases, the upper level starts being populated, degrading $g^{(2)}(0)$ as shown. {\color{blue}We note that even the larger 10 nm nanodisk with less FOM also shows a strong antibunching behavior because of the FOM value above unity.}

{ \color{blue} Finally, we investigate the non-Gaussianity of the states. The non-Gaussianity (nG) is crucial for a number of continuous-variable (CV) quantum information theory \cite{genoni2010quantifying}, such as the continuous-variable (CV) entanglement distillation protocol, CV quantum error correction, and cluster-state quantum computation. One of the measures for the non-Gaussianity of a state is the relative entropy measure \cite{genoni2008quantifying,genoni2010quantifying}
\begin{equation}
	\delta[\rho] = S(\tau) - S(\rho) = H(\sqrt{\det \sigma}) - S(\rho),
\end{equation}
where $H(x) = (x+1/2)\ln(x+1/2) - (x - 1/2) \ln (x-1/2)$, $\sigma$ is the covariance matrix of the state $\rho$ using the quadrature operators $q = (1/\sqrt{2})(a + a^\dag)$, $p = (1/i \sqrt{2})(a - a^\dag)$ with the boson annihilation operator $a$, and $S(\rho) = - \mathrm{Tr}[\rho \ln \rho]$ is the von Neumann entropy. Here, $\tau$ is a Gaussian state having the same $\langle p \rangle, \langle q \rangle$ and the same $\sigma$ as the state $\rho$. While this nG measure does not immediately reveal the usefulness of the state, (which indeed requires a careful consideration of the particular CV protocol of interest,) this nG measure clearly shows how far the state is from a Gaussian state, and hence, the state with a larger nG measure $\delta[\rho]$ has a potential to be useful.

It is known that the nG measure $\delta[\rho] = 0$ for any general Gaussian state $\rho = D(\alpha) S(\zeta) \nu(n) S^\dag(\zeta) D^\dag(\alpha)$ with arbitrary displacement operator $D(\alpha)$ and squeezing operator $S(\zeta)$ on the thermal state $\nu(n)$ with the average photon number $n$ \cite{genoni2008quantifying}. It is also known that $\delta[\rho]$ is non-negative, and zero only for a Gaussian state $\rho$ including $\rho = \ketbra{0}{0}$ (vacuum) \cite{genoni2008quantifying}. As a reference, we note that $\delta[\rho] = 1.39$ for $\rho = \ketbra{1}{1}$ (single photon Fock state). 

\begin{figure}[tb]
	\centering
	\captionsetup[subfloat]{oneside,margin={2.5cm,0cm}}
	\subfloat[4 nm nanodisk]{\includegraphics[width=0.47\textwidth]{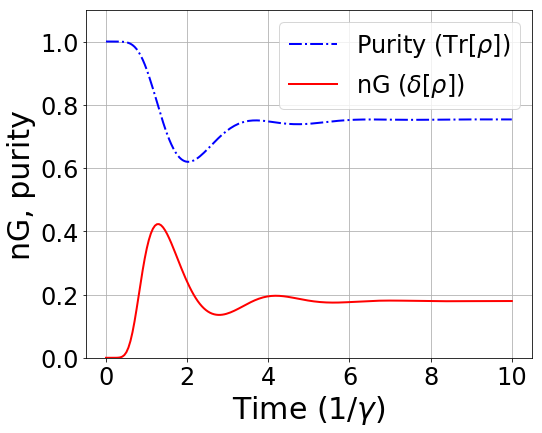}}
	\subfloat[10 nm nanodisk]{\includegraphics[width=0.47\textwidth]{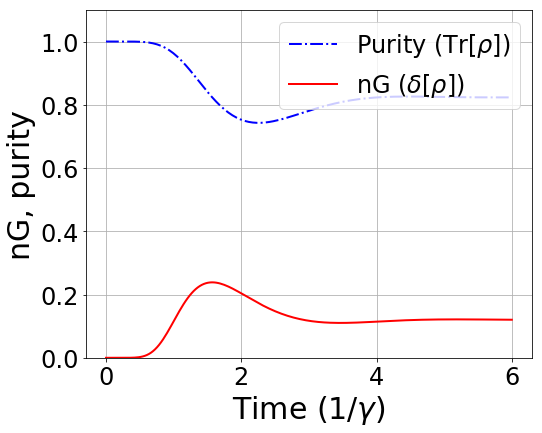}}\\
	\subfloat[4 nm nanodisk]{\includegraphics[width=0.49\textwidth]{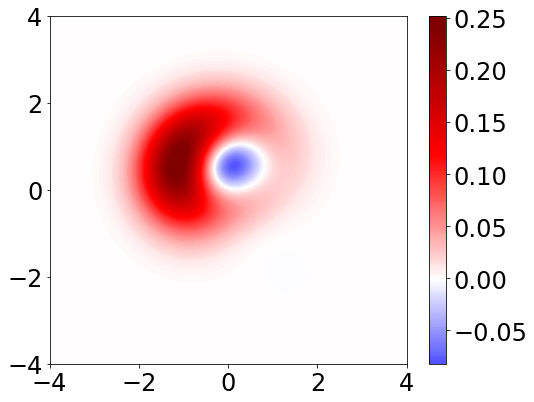}}
	\subfloat[10 nm nanodisk]{\includegraphics[width=0.47\textwidth]{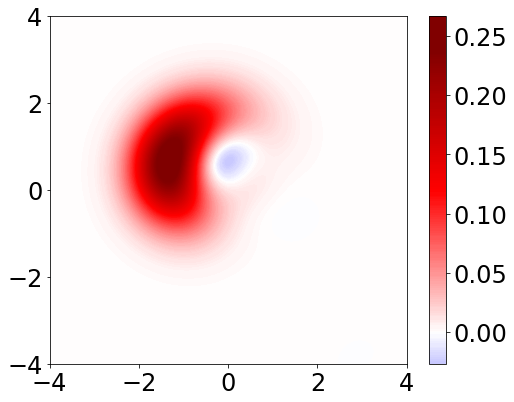}}
	\caption{Upper: transient evolution of the nG measure $
	\delta[\rho(t)]$ and the purity $\mathrm{Tr}[\rho(t)]$, lower: Wigner function of the transient states with the peak nG values, for a 4 nm nanodisk ($\hbar \chi = 53$ meV, $\hbar \chi' = 13$ meV) and a 10 nm nanodisk ($\hbar \chi = 28$ meV, $\hbar \chi' = 16$ meV).}
	\label{fig:gN}
\end{figure}

Figure \ref{fig:gN} shows the simulated results of nG measure for the 4 nm nanodisk and 10 nm nanodisk cases, respectively, as a function of time. The initial state is vacuum and the pumping starts at time $t = 0$ with the pump power of $\sqrt{\kappa} \varepsilon = 27$ meV. Both 4 nm and 10 nm nanodisks show substantial nG. Particularly, the nG peaks in the first Rabi oscillation cycle, reaching 0.42 (0.24) at time $t = 1.3/\gamma (t = 1.6/\gamma)$ for 4 nm (10 nm) nanodisk case, respectively. The transient state with the peak nG shows a strong quantum feature of negative Wigner function values, although the purity of the state degraded. The purity at the peak nG is 0.79 (0.82) for 4 nm (10 nm) nanodisks. Such purity degradation of the state is expected in such a dissipative system. Then, one can control the pump to last only up to the point of the nG peak, which will prepare the state having strong quantum signatures. Such state with negative Wigner function values and meaningful nG values may become quite useful for CV quantum information applications. 
}

\section{Conclusion and discussions}

We have demonstrated a nonlinear oscillator composed of an LSPR with added Kerr nonlinearity via coupling to monolayer MoS$_2$. To qualify as a quantum nonlinear oscillator, the nonlinear FOM $4 \chi^2/(\gamma + \chi')^2$ must be larger than unity. The nonlinearity of a state-of-the-art smallest photonic-crystal point defect cavity filled with silicon is inadequate (FOM $\sim 10^{-6}$), whereas our proposed system has predicted FOM $\sim10$, which is a seven order-of-magnitude improvement. The improvement stems from the strong field enhancement provided by a metal nanoparticle LSPR and direct coupling to a highly nonlinear monolayer TMD material, which is thin enough to overlap well with the tightly confined LSPR evanescent field. (Although graphene possesses an order-of-magnitude larger real $\chi^{(3)}$ at the same optical frequency than monolayer TMDs, the unfavorable real-to-imaginary $\chi^{(3)}$ ratio of graphene spoils the potential advantage \cite{soh2016comprehensive}, making graphene a less desirable material than the monolayer TMDs.) Moreover, by appropriately detuning the drive frequency, one can maximize the nonlinear FOM thanks to the discrete nonlinear susceptibility spectrum of the monolayer TMDs. It is possible that the practically achievable LSPR loss rate $\gamma$ could be larger than the theoretical value of 21 meV that we used in our calculation, for example, due to surface impurities, and in this case, the FOM would degrade accordingly. Even in the weakly nonlinear regime, however, pulse sequences analogous to those designed for suppression of population outside the qubit subspace in microwave anharmonic oscillators~\cite{motzoi2009simple} could be used to enable manipulation of the LSPR state in an effectively nonlinear fashion. 

Our proposed MNP-TMD quantum nonlinear oscillator could be further coupled to an optical cavity, facilitating its use to construct nonlinear quantum photonic devices. Recent studies demonstrated a decoherence-free operation of a system combining an LSPR mode and an emitter~\cite{peng2017enhancing,gurlek2017manipulation}. { \color{blue} For example, one can couple our quantum nonlinear oscillator with a cavity. In this case, one can consider the following Hamiltonian
\begin{eqnarray}
H &= H_b + H_c + H_I + H_\mathrm{drive}, \nonumber \\
H_b &= \hbar \omega_b b^\dag b + \hbar (\chi_b/2) (b^\dag)^2 b^2, \nonumber \\
H_c &= \hbar \omega_c c^\dag c, \nonumber \\
H_I &= - g (b^\dag c + b c^\dag), \nonumber \\
H_\mathrm{drive} &= -i \hbar \sqrt{\kappa} (b^\dag \varepsilon e^{- i \omega t} - b \varepsilon^* e^{i \omega t}), \label{eq:Hamiltonian}
\end{eqnarray}
where $b$, $c$ are the annihilation operators for the nonlinear oscillator mode and the cavity mode, respectively, and $g$ is the coupling coefficient between the cavity and the nonlinear oscillator. The nature of the coupling is through the electron multipole interaction between the cavity and the nonlinear oscillator. The Lindblad operators are $L_1 = \sqrt{\gamma_b} b$, the (Ohmic) dissipation of the LSPR, $L_2 = \sqrt{\chi'/2} b^2$, the two-photon absorption from the monolayer MoS$_2$, and $L_3$ = $\sqrt{\gamma_c} c$, the cavity dissipation. Although the detailed description of this advanced system is beyond the scope of the current manuscript, one can use the adjoint master equation \cite{breuer2002theory} to obtain the following Heisenberg dynamical equation for the operators:
\begin{eqnarray}
\dot{b} &= - i \Delta_b b - i \chi_b(b^\dag b) b + i g c - \frac{\gamma_b}{2} b - \frac{\chi'_b}{2} (b^\dag b) b + \sqrt{\kappa} \varepsilon, \nonumber \\
\dot{c} &= - i \Delta_c c + i g b - \frac{\gamma_c}{2} c. \label{eq:Adjoint-M}
\end{eqnarray}
It is noteworthy that, in our preliminary study of such systems, the dissipated cavity mode $\sqrt{\gamma_c} c$ inherited the non-Gaussianity and the antibunching properties of the LSPR field substantially when $\gamma_c$ is comparably large as $\gamma_b$. Then, one can extract the nonlinear quantum behavior of the LSPR efficiently into a traveling wave, which is essential for a potential large-scale quantum information processing network. }

\section*{Acknowledgment}

Sandia National Laboratories is a multimission laboratory managed and operated by National Technology and Engineering Solutions of Sandia, LLC, a wholly owned subsidiary of Honeywell International, Inc., for the DOE’s National Nuclear Security Administration under contract DE-NA0003525.

\appendix

\section{Field quantization in a dielectric cavity weakly coupled with a 2D nonlinear material} \label{sec:Ham}

We follow the treatment in Drummond and Hillery \cite{drummond2014quantum}. We will use the displacement field $\vt{D}$ instead of the electric field $\vt{E}$ and make the quantization consistent with the Maxwell equation. The Maxwell equation for a nonmagnetic material is given as
\begin{eqnarray}
&\vt{\nabla} \cdot \vt{D} = 0, \quad \vt{\nabla} \cdot \vt{B} = 0, \\
&\vt{\nabla} \times \vt{E} = - \frac{\partial \vt{B}}{\partial t}, \quad \vt{\nabla}  \times \vt{H} = \frac{\partial \vt{D}}{\partial t},
\end{eqnarray}
where $\vt{D} = \epsilon_0 \vt{E} + \vt{P}$ is the displacement field, $\vt{E}$ is the electric field, $\vt{P}$ is the polarization, $\vt{B} = \mu \vt{H}$ is the magnetic field. Here, $\epsilon_0$ is the vacuum permittivity and $\mu$ is the magnetic permeability. The linear polarization is given by the following relation:
\begin{equation}
P_j = \epsilon_0 \sum_{k=1}^3 \chi_{jk}^{(1)} E_k,
\end{equation}
where $j,k = x,y,z$ and $\chi^{(1)}_{jk}$ is the linear susceptibility tensor element of the medium. 

The nonlinear polarization is typically defined by
\begin{equation}
\vt{P} = \epsilon_0 \sum_{n>0} \vt{\chi}^{(n)} : \vt{E}^{\otimes n},
\end{equation}
where $[\vt{\chi}^{(n)}]_{jklm\cdots} = \chi^{(n)}_{jklm\cdots}$. Hence, we have
\begin{equation}
P_j = \epsilon_0 \left[ \sum_k \chi^{(1)}_{jk} E_k + \sum_{k,l} \chi^{(2)}_{jkl} E_k E_l + \sum_{k,l,m} \chi^{(3)}_{jklm} E_k E_l E_m + \cdots \right].
\end{equation}
One can express the displacement field as
\begin{equation}
\vt{D} = \sum_{n>0} \vt{\epsilon}^{(n)} : \vt{E}^{\otimes n},
\end{equation}
where $\vt{\epsilon}$ is the material's electric permittivity tensor.

We next introduce equivalent formalism based on the expansion of the electric field with respect to the polynomials of the displacement field:
\begin{equation}
\vt{E} = \sum_{n>0} \vt{\eta}^{(n)} : \vt{D}^{\otimes n},
\end{equation}
where $\vt{\eta}^{(n)}$ is the inverse permittivity tensor. The relation between $\vt{\epsilon}$ and $\vt{\eta}$ is found by equating
\begin{equation}
\vt{E} = \sum_{n>0} \vt{\eta}^{(n)} : \left[ \sum_{m>0} \vt{\epsilon}^{(m)} : \vt{E}^{\otimes m} \right]^{\otimes n}.
\end{equation}
In the case of the 2D TMD material, when we use the polarization basis such that $\hat{\varepsilon}_+, \hat{\varepsilon}_-, \hat{\varepsilon}_z$ where the first two are the clockwise and the counter-clockwise circular polarization in the x-y plane, the dielectric tensor is indeed diagonal \cite{soh2018optical}. Then, the displacement field is given by
\begin{eqnarray}
    D_+ &= \epsilon_0 \left( (1 + \chi^{(1)}) E_+ + 3\chi^{(3)} E_{+}^3 \right), \nonumber \\
    D_- &= \epsilon_0 \left( (1 + \chi^{(1)}) E_- + 3\chi^{(3)} E_{-}^3 \right), 
\end{eqnarray}
where $\chi^{(1)} = \chi^{(1)}_+ = \chi^{(1)}_-$ and $\chi^{(3)} = \chi^{(3)}_{++++} = \chi^{(3)}_{----}$. Here, the factor 3 is the degeneracy such that $\chi^{(3)} = \chi^{(3)}(\omega; \omega, \omega, -\omega) = \chi^{(3)}(\omega; \omega, -\omega, \omega) = \chi^{(3)}(\omega; -\omega, \omega, \omega)$. From this, we have
\begin{eqnarray}
\epsilon^{(1)}_{++} = \epsilon^{(1)}_{--} = \epsilon_0 (1 + \chi^{(1)}), \quad \epsilon^{(1)}_z = \epsilon_0, \quad \epsilon^{(3)}_{++++} = \epsilon^{(3)}_{----} = 3 \epsilon_0 \chi^{(3)}, 
\end{eqnarray}
while all the other elements up to the third order tensors are zero. 

The Hamiltonian is given as \cite{drummond2014quantum}
\begin{eqnarray}
H &= \int d^3 r \left[ \frac{1}{2\mu} | \vt{B}|^2 + \sum_{n \ge 1} \frac{1}{n+1} \vt{D} \cdot (\vt{\eta}^{(n)} : \vt{D}^{\otimes n}) \right] \nonumber \\
&= \int d^3 r \left[ \frac{1}{2 \mu } | \vt{B}|^2 + \sum_{n \ge 1} \frac{n}{n+1} \vt{E} \cdot (\vt{\epsilon}^{(n)} : \vt{E}^{\otimes n}) \right].
\end{eqnarray}
For a a linear medium, the quantized displacement field is obtained as \cite{drummond2014quantum}:
\begin{equation}
\vt{D} = i \epsilon (\vt{r})\sqrt{\frac{\hbar \omega }{2}} (a \vt{f} (\vt{r}) - a^\dag \vt{f}^* (\vt{r})), \label{eq:D-field}
\end{equation}
where the mode function $\vt{f}(\vt{r})$ is normalized such that
\begin{equation}
\int d^3 r ~\epsilon (\vt{r}) \vt{f}^* (\vt{r}) \cdot \vt{f} (\vt{r}) = 1. \label{eq:mode-function-norm}
\end{equation}
For example, the electric field in a 1D Fabry-Perot linear isotropic dielectric cavity is quantized through
\begin{equation}
\vt{E} = \frac{1}{\epsilon(\vt{r})} \vt{D} = i \sqrt{\frac{\hbar \omega}{2 \epsilon (\vt{r}) V}} (a e^{i \vt{k} \cdot \vt{z}} - a^\dag e^{- i \vt{k} \cdot \vt{z}}),
\end{equation}
where $V$ is the cavity volume. 

Next, we consider a case where a small piece of nonlinear material is coupled to the cavity field. When the nonlinear material is substantially occupying the cavity volume, one must follow the procedure in Drummond and Hillery \cite{drummond2014quantum} where a dual potential $\vt{\Lambda}$ such that $\vt{D} = \vt{\nabla} \times \vt{\Lambda}$ is introduced and the quantized field is not directly $\vt{D}$, but this dual potential such that
\begin{equation}
\vt{\Lambda} = \sqrt{\frac{\hbar}{2 \mu \omega}} (a \vt{u}(\vt{r}) - a^\dag \vt{u}^* (\vt{r})),
\end{equation}
where $\vt{u}(\vt{r})$ is the mode function of the dual potential and $a,a^\dag$ are the annihilation and the creation operators of the quantized field. Then, $\vt{D}$ is obtained by taking the curl of this dual potential. However, when the overlap between the cavity field and the nonlinear material is sufficiently small, one can avoid such cumbersome treatment by taking the leading order perturbing term of the Hamiltonian such that
\begin{equation}
H = H_0 + H_\mathrm{nl},
\end{equation}
where
\begin{eqnarray}
    H_0 &= \int d^3 r \left[ \frac{1}{2 \mu} | \vt{B}|^2 + \frac{1}{2} \vt{D} \cdot ( \vt{\eta}^{(1)} : \vt{D}) \right] \nonumber \\
    &= \int d^3 r \left[ \frac{1}{2 \mu} | \vt{B}|^2 + \frac{1}{2} \vt{E} \cdot ( \vt{\epsilon}^{(1)} : \vt{E}) \right], \nonumber \\
    H_\mathrm{nl} &= \int d^3 r \frac{m}{m+1} \vt{E} \cdot ( \vt{\epsilon}^{(m)} : \vt{E}^{\otimes m}), \label{eq:perturb-Hamilton}
\end{eqnarray}
where $m$ is the leading order of the perturbation. Then, the first term is converted to $H_0 = \hbar \omega a^\dag a$ using the quantized field $\vt{D}$ in \Eref{eq:D-field} and the field annihilation and creation operators.

We note that the same approach was adopted in many earlier papers regarding the Kerr-nonlinear optical cavity \cite{imamoglu1997strongly,ferretti2012single,miranowicz2013two,li2017cascaded}. In this approximate approach, $H_0$ represents the unperturbed Hamiltonian and $H_\mathrm{nl}$ is the perturbing Hamiltonian. This approach is valid when $E_\mathrm{nl} \ll E_\mathrm{0}$ where $E_i = \langle H_i \rangle$ with $i = \mathrm{0}, \mathrm{nl}$. This condition is usually met when the nonlinear 2D material's overlap with the cavity field is sufficiently small. The advantage of this approach is that the unperturbed Hamiltonian is easily quantized while the perturbing nonlinear term can be expanded using the quantized field $\vt{D}$ from the unperturbed Hamiltonian, which is given by \Eref{eq:D-field}.

\section{Nonlinear coefficient $\chi, \chi'$ and unperturbed Hamiltonian} \label{sec:chi}

The nonlinear coefficient $\chi_b$ appearing in the Hamiltonian depends on the details such as the shape, the size, and the material properties of the MNP and the nonlinear material. Once they are known, the value of $\chi_b$ for a particular LSPR mode can be accurately calculated. We solve $\chi_b$ for an arbitrary quantized field of an MNP. 

The second-quantized displacement field of the LSPR mode in an arbitrary MNP is \cite{WAKS2010CAVITY,peng2017enhancing}:
\begin{equation}
\vt{D} (\vt{r},t) = i \epsilon (\vt{r})  \sqrt{\frac{\hbar \omega_a}{2}} \vt{f} (\vt{r}) (a(t) - a^\dag(t)), \label{eq:quantized-LSPR-field}
\end{equation}
where $\omega_a$ is the resonance frequency of the LSPR, $a$ and $a^\dag$ are the annihilation and the creation operators of the LSPR with the usual bosonic commutator $[a,a^\dag] = 1$, and the mode function $\vt{f}(\vt{r})$ is normalized according to \Eref{eq:mode-function-norm}. This form is consistent with the quantized $\vt{D}$ field appearing in \Eref{eq:D-field}. 

The nonlinear optical material, namely MoS$_2$ in our case, is embedded into a background material such as the hexagonal boron nitride that fills the space. The refractive index of MoS$_2$ at 1.06 eV is calculated through DFT method \cite{ullah2018analysis} to be 2.2. The refractive index of the hexagonal boron nitride is also found to be 2.2 \cite{cappellini2001optical}. Therefore, for simplicity, we assume that the linear dielectric function of the external space of the MNP is a constant with respect to the position $\vt{r}$. We adopt the approximate perturbative approach in the previous section to consider an LSPR mode with a weakly coupled nonlinear monolayer TMD material. The perturbation is expected to be small due to the small spatial overlap of the LSPR field with the atomically thin monolayer. 

Let us now consider the nonlinear perturbation $H_\mathrm{nl}$ in \Eref{eq:perturb-Hamilton}. The leading order is the third-order nonlinearity \cite{soh2018optical} and, therefore, we obtain
\begin{equation}
H_\mathrm{nl} = \int d^3 r  \frac{9}{4} \epsilon_0 \mathrm{Re}[\chi^{(3)} (\vt{r})] |\vt{E}_\parallel |^4, \label{eq:Hnl}
\end{equation}
where $\vt{E}_\parallel$ is the electric field component parallel to the 2D surface. We also explicitly noted that $\chi^{(3)}$ is a function of space so that only the in-plane electric field component in the nonlinear 2D material contributes to the third-order optical nonlinearity. The quantized in-plane component of the electric field of the unperturbed Hamiltonian is given from the quantized unperturbed displacement field $\vt{D}$ as
\begin{equation}
\vt{E}_\parallel (\vt{r}) = i \sqrt{\frac{\hbar \omega_a}{2}} \vt{f}_\parallel (\vt{r}) (a - a^\dag),
\end{equation}
where $\vt{f}_\parallel (\vt{r})$ is the in-plane component of $\vt{f} (\vt{r})$ in the 2D material. Applying the rotating wave approximation, we calculate
\begin{equation}
(a - a^\dag)^4 = 6 (a^\dag)^2 a^2 + 12 a^\dag a + 3.
\end{equation}
We then obtain the nonlinear part
\begin{eqnarray}
H_\mathrm{mnp,Kerr} &= \frac{27 \epsilon_0 \hbar^2 \omega_a^2}{16} (2 (a^\dag)^2 a^2 + 4 a^\dag a + 1) \int d^3 r \mathrm{Re} [ \chi^{(3)} (\vt{r}) ] |\vt{f}_\parallel (\vt{r})|^4 \nonumber \\
&= E_0 + \hbar \Delta_k' a^\dag a + \hbar \frac{\chi}{2} a^\dag a^\dag a a,
\end{eqnarray}
where the first term is an uninsteresting constant and the second term is the Lamb shift from the Kerr process. Here, $\vt{f}_\parallel (\vt{r})$ is the in-plane component of the LSPR mode function. Therefore, we obtain 
\begin{equation}
\chi = \frac{27\epsilon_0 \hbar \omega_a^2}{4} \int d^3 r \mathrm{Re} [\chi^{(3)} (\vt{r}) ] |\vt{f}_\parallel (\vt{r})|^4. \label{eq:chi-sup}
\end{equation}
Since $\chi$ and $\chi'$ are related through the Kramers-Kronig relation, the above leads to the result in the main text (Eq. (3)). This is indeed a general result regardless of the details of the LSPR field distribution $\vt{f}(\vt{r})$. It also applies to any optical cavity.

It is straightforward to derive the equation (1) of the main text as follows. The quantized electric field in a 1D Fabry-Perot cavity is given from \Eref{eq:D-field} as
\begin{equation}
\vt{E} (z,t) = i\sqrt{\frac{\hbar \omega_c}{2 \epsilon V_c}} \sin (n (\omega_c/c_0) z) \hat{x} \left( c e^{- i \omega_c t} - c^\dag e^{i \omega_c t}  \right),
\end{equation}
where $\omega_c$ is the cavity's mode's frequency and $c$ is the cavity's annihilation operator. Hence, we obtain
\begin{equation}
\vt{f}(\vt{r}) = \frac{1}{\sqrt{\epsilon V_c}} \sin (n (\omega_c/c_0) z) \hat{x}.
\end{equation}
Hence, one can calculate $\chi$ using this through the \Eref{eq:chi-sup}.

Finally, we discuss the unperturbed Hamiltonian of the combined system. The nonlinear term is an addition to the linear harmonic oscillator Hamiltonian $\hbar a^\dag a$ with the annihilation operator $a$ of the quantized LSPR field \cite{WAKS2010CAVITY}, so that the unperturbed Hamiltonian is
\begin{equation}
H_a = \hbar \omega_a a^\dag a + \frac{\hbar \chi}{2} a^\dag a^\dag a a,
\end{equation}
which coincides with Eq. (1) in the main text.

\section{Population of the LSPR levels} \label{sec:pop}

To study the population of the anharmonic ladder of the LSPR, we consider the anharmonic energy eigenstates of the LSPR mode: particularly three lowest levels $\ket{0}_b, \ket{1}_b, \ket{2}_b$ whose energy eigenvalues are $0, \hbar \omega_b , \hbar(2\omega_b + \chi_b)$. An external field drives the transition among the levels. We are particularly interested in the ratio of the population at the second level to that at the first level. The dynamics of this anharmonic system can be modeled using a Hamiltonian $H' = H'_{a} + H'_\mathrm{drive}$ with
\begin{eqnarray}
H'_{a} &= \hbar \omega_a \ketbra{1}{1} + \hbar (2\omega_a + \chi) \ketbra{2}{2}, \nonumber \\
H'_\mathrm{drive} &= - i \hbar \sqrt{\kappa} \left[ (\ketbra{1}{0} + \ketbra{2}{1}) \varepsilon e^{- i \omega t} - (\ketbra{0}{1} + \ketbra{1}{2}) \varepsilon^* e^{i \omega t} \right],
\end{eqnarray}
and the Lindblad dissipation operators:
\begin{equation}
L_1 = \sqrt{\gamma}\ketbra{0}{1}, \quad L_2 = \sqrt{\gamma} \ketbra{1}{2}, \quad L_3 = \sqrt{\chi'} |0 \rangle \langle 2|.
\end{equation}

The master equations for the density matrix elements are obtained by \cite{breuer2002theory}:
\begin{equation}
\dot{\rho} = \frac{i}{\hbar} [ \rho, H] + \sum_j \left( L_j \rho L_j^\dag - \frac{1}{2} L_j^\dag L_j \rho - \frac{1}{2} \rho L_j^\dag L_j \right).
\end{equation}
For simplicity, we consider the case where $\omega = \omega_a$ (resonant drive) and $\varepsilon^* = \varepsilon$. When we separately treat the dynamics only between $\ket{1}$ and $\ket{2}$ levels and transform onto a rotating frame, the equations of motion are
\begin{eqnarray}
\dot{\rho}_{22} &= - \sqrt{\kappa} \varepsilon (\rho_{21} + \rho_{12} ) - (\gamma + \chi') \rho_{22}, \nonumber \\
\dot{\rho}_{21} &= - i\chi \rho_{21} + \sqrt{\kappa} \varepsilon (\rho_{22} - \rho_{11})  - \frac{\gamma + \chi'}{2} \rho_{21}, \nonumber \\
\dot{\rho}_{12} &=  +i \chi \rho_{12} + \sqrt{\kappa} \varepsilon(\rho_{22} - \rho_{11}) - \frac{\gamma + \chi'}{2} \rho_{12},
\end{eqnarray}
where we used the fact $\rho_{20} = 0$. Here, we denoted $\rho_{ij} = \bra{i} \rho \ket{j}$. One can solve the above equation using the Fourier transformation technique:
\begin{eqnarray}
- i \omega' \tilde{\rho}_{22} (\omega') &= - \sqrt{\kappa}\varepsilon (\tilde{\rho}_{21} (\omega') + \tilde{\rho}_{12} (\omega')) - (\gamma + \chi') \tilde{\rho}_{22}, \nonumber \\
- i \omega' \tilde{\rho}_{21} (\omega') &= - i \chi \tilde{\rho}_{21} (\omega') + \sqrt{\kappa} \varepsilon (\tilde{\rho}_{22} (\omega') - \tilde{\rho}_{11} (\omega')) - \frac{\gamma + \chi'}{2} \tilde{\rho}_{21}, \nonumber \\
- i \omega' \tilde{\rho}_{12} (\omega') &= + i \chi \tilde{\rho}_{12} (\omega') + \sqrt{\kappa} \varepsilon (\tilde{\rho}_{22} (\omega') - \tilde{\rho}_{11} (\omega')) - \frac{\gamma + \chi'}{2} \tilde{\rho}_{12},
\end{eqnarray}
From the second and the third equations, we obtain
\begin{eqnarray}
\tilde{\rho}_{21} (\omega') &= \frac{\sqrt{\kappa} \varepsilon (\tilde{\rho}_{22} (\omega') - \tilde{\rho}_{11} (\omega'))}{i ( \chi - \omega') + (\gamma + \chi')/2}, \nonumber \\
\tilde{\rho}_{12} (\omega') &= \frac{\sqrt{\kappa} \varepsilon (\tilde{\rho}_{22} (\omega') - \tilde{\rho}_{11} (\omega'))}{-i ( \chi + \omega') + (\gamma + \chi')/2}.
\end{eqnarray}
From this, we obtain the steady-state solution by setting $\omega' \rightarrow 0$:
\begin{eqnarray}
\tilde{\rho}_{22} (\omega' \rightarrow 0) = \frac{4 \kappa \varepsilon^2 }{(\gamma + \chi')^2 + 4 \chi^2 + 4 \kappa \varepsilon^2} \tilde{\rho}_{11}.
\end{eqnarray}

When $\kappa \varepsilon^2$ is large, the population in $\ket{2}$ becomes saturated. For a sufficiently small drive $\varepsilon$ such that $4 \kappa \varepsilon^2 \ll {\gamma}^2$, we obtain the steady-state population
\begin{equation}
\rho_\mathrm{22,ss} \simeq \frac{4 \kappa \varepsilon^2}{(\gamma + \chi')^2 (1 + 4 \chi^2/(\gamma + \chi')^2)} \rho_\mathrm{11,ss}. 
\end{equation}

\section{An example design of a metal nanoparticle} \label{sec:design}

In this section, we illustrate one possible design of the metal nanoparticle and the associated LSPR mode that provides a sufficiently large Kerr coefficient in the Hamiltonian at our desired driving optical frequency. We consider a configuration in the figure 1 of the main text where an LSPR field is coupled to the two 2D material planes in the left and the right. One drives the LSPR resonance using a classical light source (lasers) as shown.

\begin{figure}
	\centering
	\includegraphics[width=0.45\textwidth]{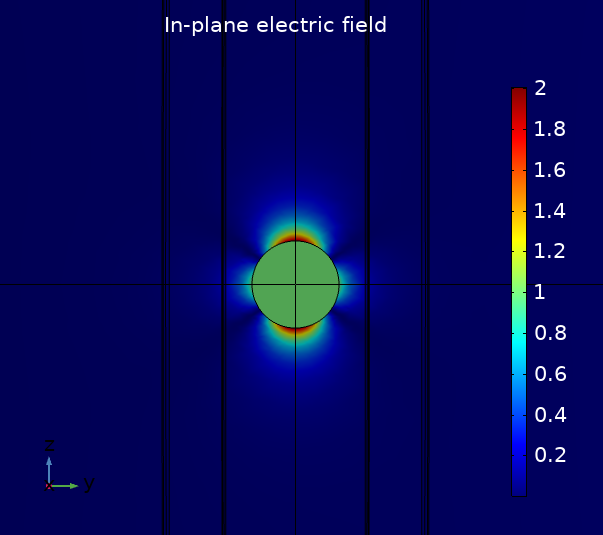}
	\caption{In-plane component of the LSPR field in a spherical MNP. Also shown are the two monolayer MoS$_2$ planes in the left and the right. (Solved in COMSOL FEM software package)}
	\label{fig:spherical_MNP}
\end{figure}

The mode field function and the resonance frequency for the spherical MNPs are well known (for example, see \cite{shvets2012plasmonics}). Unfortunately, however, the spherical MNP does not provide a large overlap between the in-plane LSPR field component and the sandwiching 2D material planes: To excite the in-plane field component, the electrical field component of the driving field must be parallel to the 2D planes. Then, most of the in-plane field component of spherical LSPR is concentrated in the middle of the sphere \cite{WAKS2010CAVITY} (also see the FEM solution in the figure \ref{fig:spherical_MNP}). Therefore, we must find another geometry of the MNP that provides a large overlap. It is also well known that the detailed LSPR field function (distribution) and the resonance depend largely on the shape and size of the MNP \cite{link2000shape,kelly2003optical}. A reasonable path forward is to utilize the finite element method to solve the Maxwell equations in details. Hence, we adopt the commercial FEM software package (COMSOL) to design a suitable MNP that provides the resonance peak at the desired optical frequency at 1.06 eV and maximizes the nonlinear overlap.

\begin{figure}[!tb]
	\centering
	\subfloat[Side]{\includegraphics[width=0.42\textwidth]{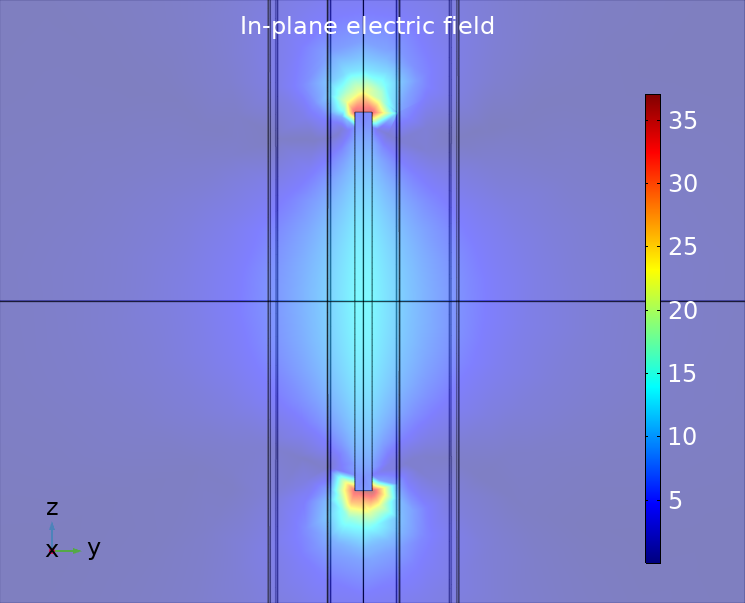}}\quad
	\subfloat[Front]{\includegraphics[width=0.42\textwidth]{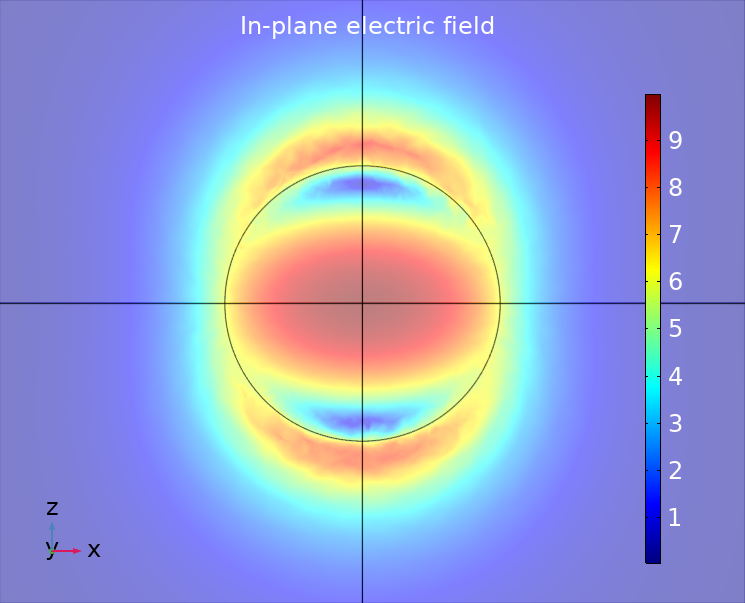}} \\
	\subfloat[E-field vectors]{\includegraphics[width=0.6\textwidth]{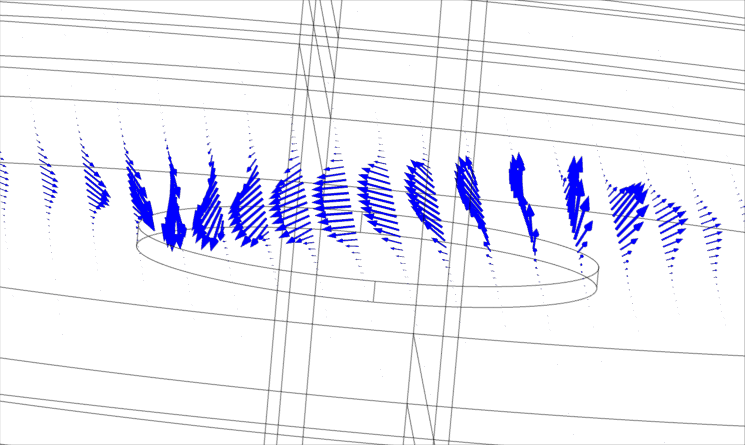}}
	\caption{E-field distribution of a silver disk MNP driven by an external field. Solution obtained using a finite-element-method software package (COMSOL).}
	\label{fig:disk-MNP}
\end{figure}

A natural selection of the MNP geometry that may improve the in-plane component overlapping the nonlinear 2D material is a disk shape; the large bottom and top surfaces induce large in-plane fields, while the thin layer of the disk may help make the resonant frequency lower. For example, we solved the field distribution of an LSPR mode from a silver disk MNP with diameter 6 nm and thickness 0.27 nm at 1140 nm wavelength, which is the resonant frequency of this MNP design. For this, we applied a $z$-polarized external field, traveling in $y-$ direction with 1 V/m amplitude to the system and obtained the induced electric field (subtracting the background). Figure \ref{fig:disk-MNP} shows the result. The two monolayers of MoS$_2$ overlap well with the in-plane component of the LSPR induced field. The E-field vectors in the monolayer planes are significant, compared to the case of the spherical MNP.

\section*{References}
\bibliography{iopart-num}

\providecommand{\newblock}{}
\begin{thebibliography}{10}
\expandafter\ifx\csname url\endcsname\relax
  \def\url#1{{\tt #1}}\fi
\expandafter\ifx\csname urlprefix\endcsname\relax\def\urlprefix{URL }\fi
\providecommand{\href}[2]{#1}  
\providecommand{\eprint}[2][arXiv]{#1:\linebreak[0]#2}

\bibitem{mabuchi2012qubit}
Mabuchi H 2012 {\em Physical Review A\/} {\bf 85} 015806

\bibitem{imamoglu1997strongly}
Imamoglu A, Schmidt H, Woods G and Deutsch M 1997 {\em Physical Review
  Letters\/} {\bf 79} 1467

\bibitem{sun2016optical}
Sun Z, Martinez A and Wang F 2016 {\em Nature Photonics\/} {\bf 10} 227

\bibitem{autere2018nonlinear}
Autere A, Jussila H, Dai Y, Wang Y, Lipsanen H and Sun Z 2018 {\em Advanced
  Materials\/} {\bf 30} 1705963

\bibitem{soh2016comprehensive}
Soh D~B, Hamerly R and Mabuchi H 2016 {\em Physical Review A\/} {\bf 94} 023845

\bibitem{soh2018optical}
Soh D~B, Rogers C, Gray D~J, Chatterjee E and Mabuchi H 2018 {\em Physical
  Review B\/} {\bf 97} 165111

\bibitem{nielsen2017giant}
Nielsen M~P, Shi X, Dichtl P, Maier S~A and Oulton R~F 2017 {\em Science\/}
  {\bf 358} 1179--1181

\bibitem{WAKS2010CAVITY}
Waks E and Sridharan D 2010 {\em Physical Review A\/} {\bf 82} 043845

\bibitem{shvets2012plasmonics}
Shvets G and Tsukerman I 2012 {\em Plasmonics and Plasmonic Metamaterials:
  Analysis and Applications\/} vol~4 (World Scientific)

\bibitem{jensen2000nanosphere}
Jensen T~R, Malinsky M~D, Haynes C~L and Van~Duyne R~P 2000 {\em The Journal of
  Physical Chemistry B\/} {\bf 104} 10549--10556

\bibitem{miranowicz2013two}
Miranowicz A, Paprzycka M, Liu Y~x, Bajer J and Nori F 2013 {\em Physical
  Review A\/} {\bf 87} 023809

\bibitem{lin2007dispersion}
Lin Q, Zhang J, Piredda G, Boyd R~W, Fauchet P~M and Agrawal G~P 2007 {\em
  Applied physics letters\/} {\bf 91} 021111

\bibitem{jana2001wet}
Jana N~R, Gearheart L and Murphy C~J 2001 {\em The Journal of Physical
  Chemistry B\/} {\bf 105} 4065--4067

\bibitem{rogers2018laser}
Rogers C, Gray D, Bogdanowicz N and Mabuchi H 2018 {\em Physical Review
  Materials\/} {\bf 2} 094003

\bibitem{novoselov20162d}
Novoselov K, Mishchenko A, Carvalho A and Neto A~C 2016 {\em Science\/} {\bf
  353} aac9439

\bibitem{lukovsevivcius2009reservoir}
Luko{\v{s}}evi{\v{c}}ius M and Jaeger H 2009 {\em Computer Science Review\/}
  {\bf 3} 127--149

\bibitem{breuer2002theory}
Breuer H~P, Petruccione F {\em et~al\/} 2002 {\em The theory of open quantum
  systems\/} (Oxford University Press on Demand)

\bibitem{carmichael2009statistical}
Carmichael H~J 2009 {\em Statistical methods in quantum optics 2: Non-classical
  fields\/} (Springer Science \& Business Media)

\bibitem{peng2017enhancing}
Peng P, Liu Y~C, Xu D, Cao Q~T, Lu G, Gong Q, Xiao Y~F {\em et~al\/} 2017 {\em
  Physical review letters\/} {\bf 119} 233901

\bibitem{ferretti2012single}
Ferretti S and Gerace D 2012 {\em Physical Review B\/} {\bf 85} 033303

\bibitem{drummond2014quantum}
Drummond P~D and Hillery M 2014 {\em The quantum theory of nonlinear optics\/}
  (Cambridge University Press)

\bibitem{li2017cascaded}
Li A, Zhou Y and Wang X~B 2017 {\em Scientific reports\/} {\bf 7} 7309

\bibitem{ullah2018analysis}
Ullah M~S, Yousuf A~H~B, Es-Sakhi A~D and Chowdhury M~H 2018 Analysis of
  optical and electronic properties of mos2 for optoelectronics and fet
  applications {\em AIP Conference Proceedings\/} vol 1957 (AIP Publishing) p
  020001

\bibitem{cappellini2001optical}
Cappellini G, Satta G, Palummo M and Onida G 2001 {\em Physical Review B\/}
  {\bf 64} 035104

\bibitem{wang2017plane}
Wang G, Robert C, Glazov M, Cadiz F, Courtade E, Amand T, Lagarde D, Taniguchi
  T, Watanabe K, Urbaszek B {\em et~al\/} 2017 {\em Physical review letters\/}
  {\bf 119} 047401

\bibitem{echeverry2016splitting}
Echeverry J, Urbaszek B, Amand T, Marie X and Gerber I 2016 {\em Physical
  Review B\/} {\bf 93} 121107

\bibitem{kuramochi2017ultrahigh}
Kuramochi E, Kim J~K, Taniyama H, Shinya A, Kita S and Notomi M 2017
  Ultrahigh-q/v single point-defect photonic crystal nanocavity with embedded
  sub-wavelength air-slot {\em CLEO: Science and Innovations\/} (Optical
  Society of America) pp JTh3M--5

\bibitem{hales2018third}
Hales J~M, Chi S~H, Allen T, Benis S, Munera N, Perry J~W, McMorrow D, Hagan
  D~J and Van~Stryland E~W 2018 Third-order nonlinear optical coefficients of
  si and gaas in the near-infrared spectral region {\em CLEO: Applications and
  Technology\/} (Optical Society of America) pp JTu2A--59

\bibitem{yoshida1998intersubband}
Yoshida H, Mozume T, Nishimura T and Wada O 1998 {\em Electronics Letters\/}
  {\bf 34} 913--915

\bibitem{akiyama2001nonlinearity}
Akiyama T, Georgiev N, Mozume T, Yoshida H, Gopal A~V and Wada O 2001 {\em
  Electronics Letters\/} {\bf 37} 129--130

\bibitem{zakery2003optical}
Zakery A and Elliott S 2003 {\em Journal of Non-Crystalline Solids\/} {\bf 330}
  1--12

\bibitem{link2000shape}
Link S and El-Sayed M~A 2000 {\em International reviews in physical
  chemistry\/} {\bf 19} 409--453

\bibitem{kelly2003optical}
Kelly K~L, Coronado E, Zhao L~L and Schatz G~C 2003 The optical properties of
  metal nanoparticles: the influence of size, shape, and dielectric environment

\bibitem{wang2006general}
Wang F and Shen Y~R 2006 {\em Physical review letters\/} {\bf 97} 206806

\bibitem{kirakosyan2016surface}
Kirakosyan A~S, Stockman M~I and Shahbazyan T~V 2016 {\em Physical Review B\/}
  {\bf 94} 155429

\bibitem{marder2010condensed}
Marder M~P 2010 {\em Condensed matter physics\/} (John Wiley \& Sons)

\bibitem{liu2016van}
Liu Y, Weiss N~O, Duan X, Cheng H~C, Huang Y and Duan X 2016 {\em Nature
  Reviews Materials\/} {\bf 1} 16042

\bibitem{man2016protecting}
Man M~K, Deckoff-Jones S, Winchester A, Shi G, Gupta G, Mohite A~D, Kar S,
  Kioupakis E, Talapatra S and Dani K~M 2016 {\em Scientific reports\/} {\bf 6}
  20890

\bibitem{golla2013optical}
Golla D, Chattrakun K, Watanabe K, Taniguchi T, LeRoy B~J and Sandhu A 2013
  {\em Applied Physics Letters\/} {\bf 102} 161906

\bibitem{usukura2014highly}
Usukura E, Shinohara S, Okamoto K, Lim J, Char K and Tamada K 2014 {\em Applied
  physics letters\/} {\bf 104} 121906

\bibitem{tanaka2015characteristics}
Tanaka D, Imazu K, Sung J, Park C, Okamoto K and Tamada K 2015 {\em
  Nanoscale\/} {\bf 7} 15310--15320

\bibitem{masuda2017high}
Masuda S, Yanase Y, Usukura E, Ryuzaki S, Wang P, Okamoto K, Kuboki T, Kidoaki
  S and Tamada K 2017 {\em Scientific reports\/} {\bf 7} 3720

\bibitem{walls2007quantum}
Walls D~F and Milburn G~J 2007 {\em Quantum optics\/} (Springer Science \&
  Business Media)

\bibitem{kimble1992structure}
Kimble H, Polzik E, Rempe G and Thompson R 1992 Structure and dynamics in
  cavity quantum electrodynamics {\em Quantum Electronics and Laser Science
  Conference\/} (Optical Society of America) p QTuA2

\bibitem{genoni2010quantifying}
Genoni M~G and Paris M~G 2010 {\em Physical Review A\/} {\bf 82} 052341

\bibitem{genoni2008quantifying}
Genoni M~G, Paris M~G and Banaszek K 2008 {\em Physical Review A\/} {\bf 78}
  060303

\bibitem{motzoi2009simple}
Motzoi F, Gambetta J~M, Rebentrost P and Wilhelm F~K 2009 {\em Physical review
  letters\/} {\bf 103} 110501

\bibitem{gurlek2017manipulation}
Gurlek B, Sandoghdar V and Mart{\'\i}n-Cano D 2017 {\em ACS Photonics\/} {\bf
  5} 456--461

\end{thebibliography}

\end{document}